\documentclass[useAMS,usenatbib]{mn2e}     
\usepackage{times}

\usepackage{./aas_macros}
\usepackage{graphicx}
\graphicspath{{images/}}

\begin{document}
\title{Magneto-elastic oscillations and the damping of crustal shear modes in magnetars}
\author[Michael Gabler, Pablo Cerd\'a-Dur\'an, Jos\'e
A.~Font, Ewald M\"uller and Nikolaos Stergioulas ]
{Michael Gabler$^{1,2,3}$, %\thanks{E-mail:miga@mpa-garching.mpg.de}, 
Pablo Cerd\'a-Dur\'an$^1$, %\thanks{E-mail:cerda@mpa-garching.mpg.de},
Jos\'e A.~Font$^2$, %\thanks{E-mail:j.antonio.font@uv.es}
Ewald M\"uller$^1$ %\thanks{E-mail:emueller@mpa-garching.mpg.de}
\and and Nikolaos Stergioulas$^3$ %\thanks{E-mail:niksterg@auth.gr}
\\
  $^1$Max-Planck-Institut f\"ur Astrophysik,
  Karl-Schwarzschild-Str.~1, 85741 Garching, Germany \\ 
  $^2$Departamento de Astronom\'{\i}a y Astrof\'{\i}sica,
  Universidad de Valencia, 46100 Burjassot (Valencia), Spain\\
  $^3$Department of Physics, Aristotle University of Thessaloniki,
  Thessaloniki 54124, Greece }
\date{\today}
\maketitle
\begin{abstract}
In a realistic model of magneto-elastic oscillations in magnetars, we find
that crustal shear oscillations, often invoked as an explanation of 
quasi-periodic oscillations (QPOs) seen after giant flares in soft gamma-ray 
repeaters (SGRs), are damped by resonant absorption on timescales of at 
most 0.2s, for a lower limit on the dipole magnetic field strength of 
$5\times 10^{13}$G. At higher magnetic field strengths (typical in magnetars)
the damping timescale is even shorter, as anticipated by earlier toy-models. 
We have investigated a range of equations of state and masses and if magnetars 
are dominated by a dipole magnetic field, our findings exclude torsional 
shear oscillations of the crust from explaining the observed low-frequency QPOs. 
In contrast, we find that the Alfv\'en QPO model is a viable explanation of 
observed QPOs, if the dipole magnetic field strength exceeds a minimum 
strength of about several times $10^{14}$G to $10^{15}$G. Then, Alfv\'en QPOs 
are no longer confined to the fluid core, but completely dominate in the crust region 
and have a maximum amplitude at the surface of the star.
\end{abstract}
%
%=============================================================================
\section{Introduction}
The observation of giant flares in soft gamma-ray repeaters (SGRs; compact
objects with very strong magnetic fields or magnetars
\citep{Duncan1992}) may open a gateway towards the exciting field of
neutron star seismology. In the decaying X-ray tail of two such
events, SGR 1900+14 and SGR 1806-20, a number of long-lasting,
quasi-periodic oscillations (QPOs) have been observed (see
\cite{Israel2005} and \cite{Watts2007} for recent reviews).  Early
models interpreted the observed QPO frequencies as directly related to
torsional shear oscillations of the solid crust of a neutron star
excited during a giant flare event (see \cite{Duncan1998,
  Strohmayer2005, Piro2005, Sotani2007, Samuelsson2007}, and references therein) 
raising hopes that through their identification, the 
physical properties of the crust could be probed \citep{Steiner2009}.  
Due to the extremely strong magnetic fields present in magnetars, however, 
a self-consistent model that includes
global magnetohydrodynamic (MHD) oscillations interacting with the
shear oscillations of the crust is required (\cite{Levin2006},
\cite{Glampedakis2006}, \cite{Levin2007}, \cite{Lee2007,
  Lee2008}). In a highly simplified model, \cite{Levin2007} showed that
shear oscillations will be absorbed by an MHD continuum of Alfv\'en
oscillations, while long-lived QPOs may still appear at the turning
points or edges of the continuum.

\cite{Sotani2008} and subsequently \cite{Cerda2009} (see also
\cite{Colaiuda2009}), using a more realistic, general-relativistic MHD
model but still ignoring an extended crust, found two families of Alfv\'en QPOs
related to turning points of the frequency of torsional Alfv\'en waves
near the magnetic pole and inside a region of closed magnetic field
lines near the equator. Each QPO family consists of two sub-families
differing by their symmetry behaviour with respect to the equatorial
plane.  The results of the numerical simulations
were explained by a semi-analytic model based on standing waves in the
short-wavelength limit \citep{Cerda2009}. The Alfv\'en QPO model is very 
attractive, because it
reproduces the near-integer-ratios of the observed 30, 92 and 150\,Hz
frequencies in SGR 1806-20, at magnetic field strengths expected for 
magnetars. In this model, the observed SGR QPOs, rather than probing
the crust, yield information on the magnetic field and the compactness of
the star. 

The omission of an extended crust in the previous studies of
\cite{Sotani2008}, \cite{Cerda2009}, and \cite{Colaiuda2009} can be
considered as a limiting case of a very strong magnetic field.  For
intermediate magnetic field strengths, however, an understanding of
magnetar oscillations requires the inclusion of crust-core coupling.
In this Letter, we present the first such realistic simulations of coupled,
magneto-elastic oscillations. We use a general-relativistic framework,
a dipolar magnetic field, and a tabulated equation of state (EOS) for
dense matter. The numerical simulations are based on state-of-the-art
Riemann solver methods for both the interior MHD fluid and the crust.
A recent study by \cite{vanHoven10} also takes entanglement of magnetic
field lines into account, thereby generalising the toy model of
\cite{Levin2007}. Some first results on coupled crust-core oscillations also
appeared in \cite{Kokkotas2010}.

We use units where $c=G=1$ with $c$ and $G$ being the speed of light
and the gravitational constant, respectively. Latin (Greek) indices
run from 1 to 3 (0 to 3).
\vspace{-0.5cm}
%
%=============================================================================
\section{Theoretical framework}
The present study of torsional oscillations of magnetars is based on a
numerical integration of the general relativistic MHD equations. As in 
\citet{Cerda2009}, who considered purely Alfv\'en oscillations of the
fluid core, assume (i) a zero temperature
EOS, (ii) axisymmetry, (iii) a purely poloidal magnetic field
configuration, (iv) the Cowling approximation, (v) a spherically
symmetric background, and (vi) small amplitude oscillations. Because 
of assumptions (ii) and (iii) polar oscillations decouple from axial ones 
in the linear regime. Therefore, we concentrate on purely axial 
oscillations and evolve the $\varphi$-component of the evolution 
variables only. We assume a conformally flat metric 
\footnote{This provides a very good approximation as our neutron star
  models are almost perfectly spherically symmetric except for very
  small deviations due to the presence of an axisymmetric magnetic
  field.}
\begin{equation}
 ds^2 = - \alpha^2 dt^2 + \phi^4 \left( dr^2 + r^2 d\theta^2 
        + r^2 \sin{\theta}^2 d\varphi^2 \right) \, ,
\end{equation}
where $\alpha$ is the lapse function and $\phi$ the conformal factor, 
and consider a stress-energy tensor $T^{\mu\nu}$ of the form
\begin{eqnarray}
 T^{\mu\nu} &=& T^{\mu\nu}_{\mathrm{fluid}} + T^{\mu\nu}_{\mathrm{mag}} +
 T^{\mu\nu}_{\mathrm{elas}} \nonumber\\
 &=& \rho h u^\mu u^\nu + P g^{\mu\nu} +  b^2 u^\mu u^\nu + \frac{1}{2} b^2
     g^{\mu\nu} - b^\mu b^\nu \nonumber\\
 & &  - 2 \mu_{\mathrm{S}} \Sigma^{\mu\nu}\label{T^munu} \, ,
\end{eqnarray}
where $\rho$ is the rest-mass density, $h$ the specific enthalpy, $P$
the isotropic fluid pressure, $u^\mu$ the 4-velocity of the fluid,
$b^\mu$ the magnetic field measured by a co-moving observer (with
$b^2:=b^\mu b_\mu$), $\Sigma^{\mu\nu}$ the shear tensor, and
$\mu_{\mathrm{S}}$ the shear modulus. The latter is
obtained according to \citet{Sotani2007}.

The conservation of energy and momentum $\nabla_\nu T^{\mu\nu} = 0$,
and the induction equation lead to the following system of evolution
equations
\begin{equation}
 \frac{1}{\sqrt{-g}} \left( \frac{\partial\sqrt{\gamma} \bf U }{\partial t} +
 \frac{\partial \sqrt{-g} {\bf F}^i}{\partial x^i} \right) = 0 \, , 
\label{conservationlaw}
\end{equation}
where $g$ and $\gamma$ are the determinants of the 4-metric and 3-metric,
respectively. The two-component state and flux vectors are given by
\begin{eqnarray}
 {\bf U}   &=& [S_\varphi ,\, B^\varphi]  
\label{reduced_withcrust1} \, ,
 \\
 {\bf F}^r &=& \left[ - \frac{b_\varphi B^r}{W} - 2 \mu_S
                      \Sigma^r_{~\varphi} ,\, - v^\varphi B^r
               \right]  
\label{flux_r} \,  ,
\\
 {\bf F}^\theta &=& \left[ - \frac{b_\varphi B^\theta}{W}- 2 \mu_S
                          \Sigma^\theta_{~\varphi} ,\, -v^\varphi
                          B^\theta 
                   \right] \, ,
\label{flux_theta}  
\end{eqnarray}
where $B^i$ are the magnetic field components as measured by an {\it
  Eulerian observer} \citep{Anton2006}, and $W=\alpha u^t$ is the
Lorentz factor.  The shear tensor $\Sigma^{i \varphi}= 1/2 g^{ii}
\xi^\varphi_{~,i}$ contains the spatial derivatives (denoted by a
comma) of the fluid displacement $\xi^\varphi$ due to the
oscillations, which are related to the fluid 4-velocity according to $
\xi^\varphi_{~,t} =\alpha v^\varphi = {u^\varphi} / {u^t}$, where
$v^\varphi$ is the $\varphi$-component of the fluid 3-velocity. Hence,
the evolution of the spatial derivatives $\xi^\varphi_{~,r}$ and
$\xi^\varphi_{~,\theta}$ is given by
\begin{equation} 
 (\xi^\varphi_{~,k})_{,t} - ( \alpha v^\varphi)_{,k} = 0 
  \qquad \mathrm{with} \quad   k \in \{r,\theta\} \, .
\label{evo_xi}
\end{equation}

We also need to provide boundary conditions. At the surface the radial
derivative of the displacement has to vanish ($\xi^\varphi_{~,r}=0$),
as we assume a continuous traction and vanishing surface currents.  At
the crust-core interface we demand the continuity of the parallel
electric field, which implies a continuous displacement $\xi^\varphi$.
Together with the continuity of the traction the latter leads to a
relation between the radial derivatives of the displacement in the
core and in the crust: $\xi^\varphi_{\mathrm{core},r} = \left( 1 +
\delta \right) \xi^\varphi_{\mathrm{crust},r}$ with $\delta =
\mu_\mathrm{S}/(b_r b^r)$.

To construct equilibrium models we choose between different barotropic
EOSs for the core that are matched to an EOS 
for the crust. The available models for the core are the
soft EOS A \citep{Pandharipande1971}, the intermediate EOS WFF3
\citep{Wiringa1988}, EOS APR \citep{Akmal1998} and the stiff EOS L
\citep{Pandharipande1975}. For the low density region of the crust we
choose between EOS NV \citep{Negele1973} and EOS DH
\citep{Douchin2001}. Details about the different combinations of EOSs
can be found in \cite{Sotani2007}. We use a reference model with a
mass of $1.4\,M_{\odot}$ and a circumferential radius $R_\mathrm{star}
= 12.26\,$km, described by the APR$\,+\,$DH EOS. In contrast to
\cite{Sotani2007} we compute magnetised equilibrium models using the
{\small LORENE} library ({\tt www.lorene.obspm.fr}).

Our simulation code is an extended version of the GRMHD code presented
in \citet{Cerda2009}. It includes the shear terms as they appear in
(\ref{T^munu}), and the evolution of the displacements (\ref{evo_xi}).
The proper working of the MHD part of the code was demonstrated in
\citet{Cerda2008}. To test the extended code we compared its results
obtained for two limiting cases, zero magnetic field and zero shear
modulus, with those of previous studies. The purely crustal shear
oscillations presented in \citet{Sotani2007} are recovered with an
agreement of 1\%, and the Alfv\'en continuum is obtained naturally as
in \citet{Cerda2009}. Further details on the derivation of the model
equations, the numerical methods, and the code tests will be
discussed in Gabler, Cerd\'a-Dur\'an, Font \& Stergioulas (in preparation).

\vspace{-0.5cm}
%=============================================================================
\section{Damping of crustal shear modes}

To study the behaviour of coupled crust-core oscillations we perturb
the equilibrium stellar model by imposing a velocity perturbation and
then follow the time evolution of the system (\ref{conservationlaw}) -
(\ref{evo_xi}). Unless stated otherwise, we use a grid of
150$\times$100 zones in our simulations covering a domain $[0,
  R_{\mathrm{star}}] \times [0,\pi]$. The angular grid is equidistant,
while the radial grid is equidistant only in the crust, where 40 per
cent of the zones are located, and coarsens towards the centre of the
star. Symmetries are exploited whenever a perturbation is of purely
odd or even parity with respect to the equatorial plane.  We use the
term \emph{damping} in the following to refer to {\it resonant absorption}
of crustal shear oscillations by the {\it Alfv\'en continuum} of the core
(unless we explicitly refer to numerical damping caused by
finite-differencing).

%-------------------------------------------------------------------------------
%
\begin{figure*}
 \centering
 \includegraphics[width=.31\textwidth]{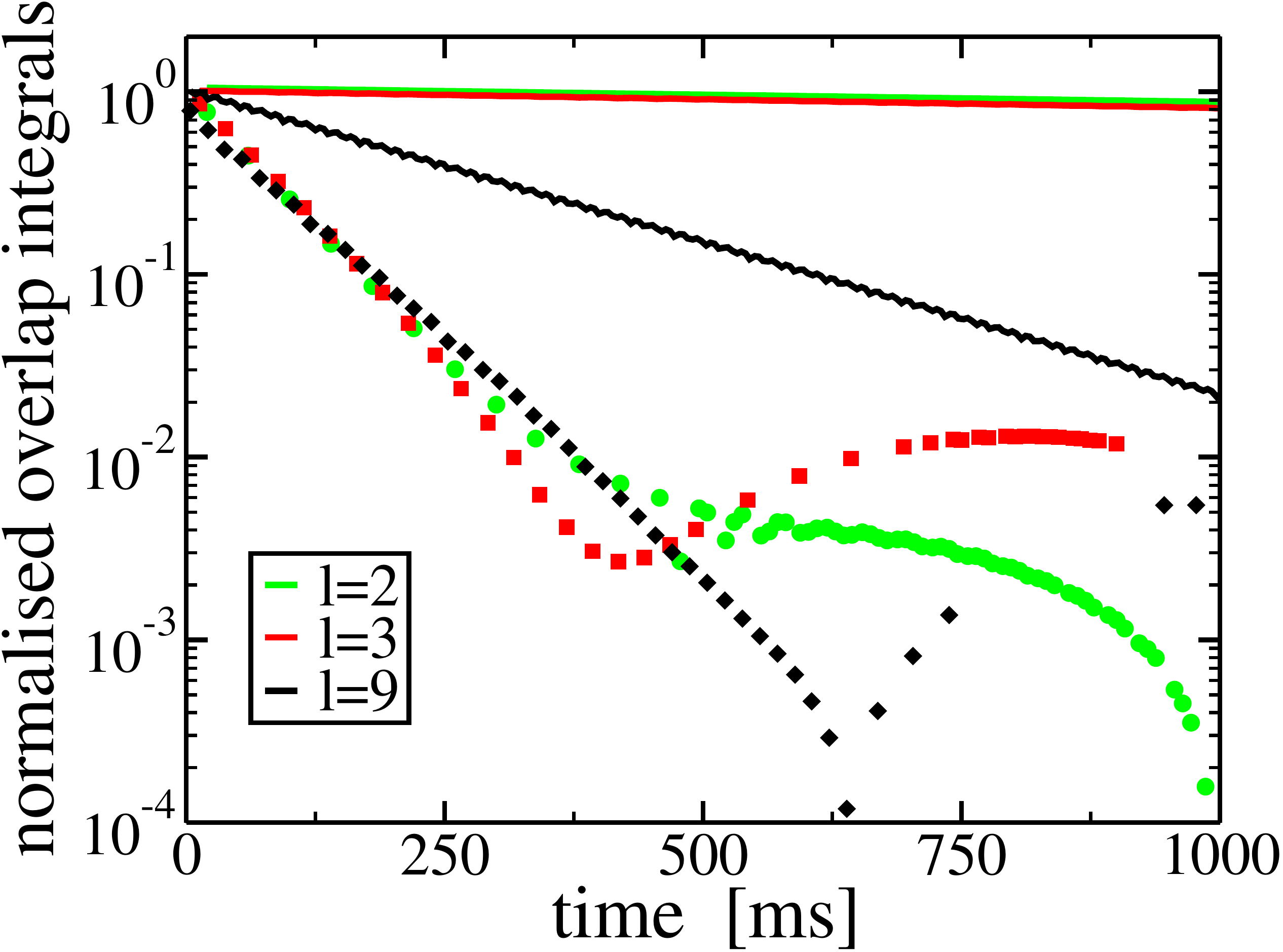} 
 \includegraphics[width=.31\textwidth]{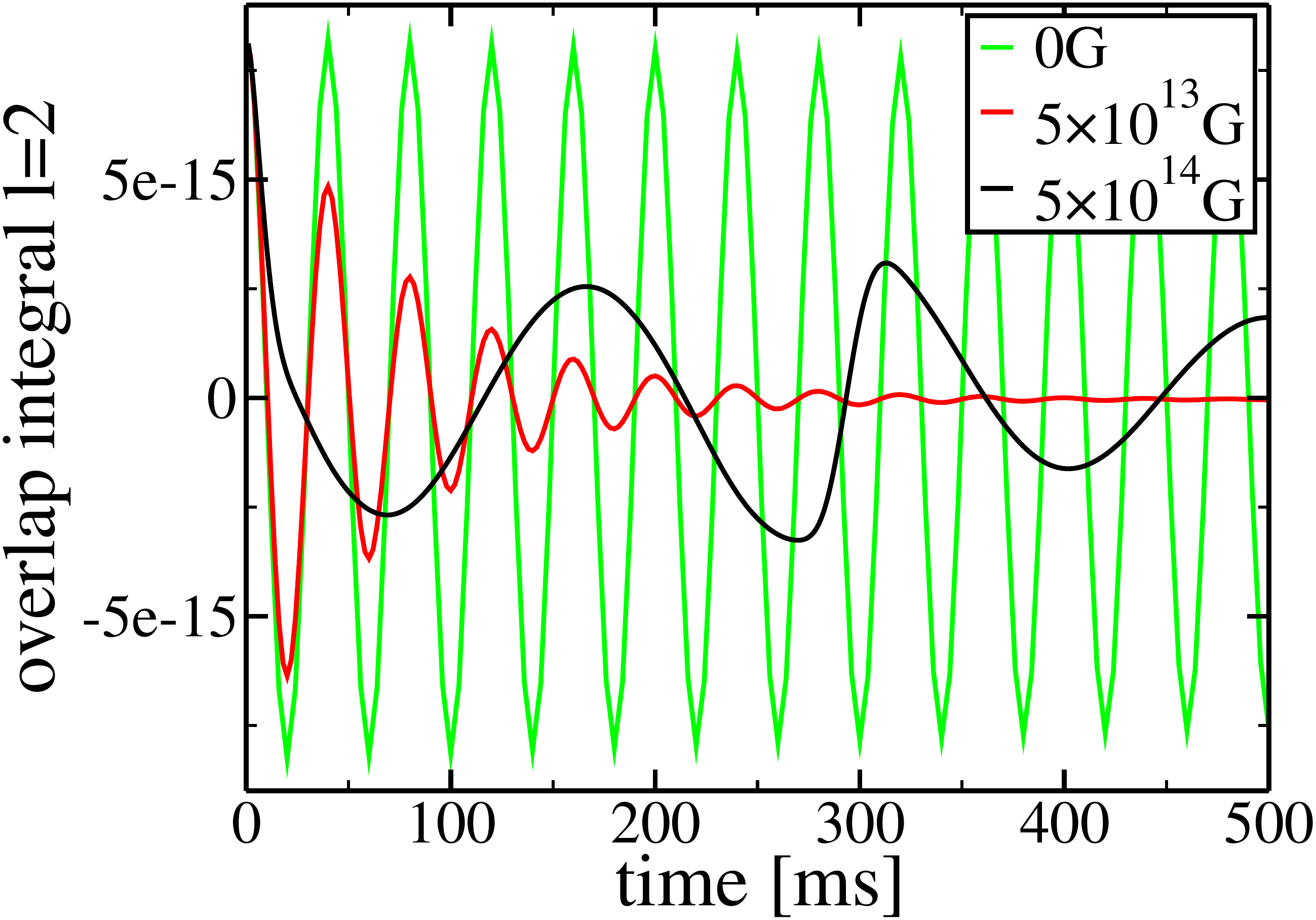}
 \includegraphics[width=.31\textwidth]{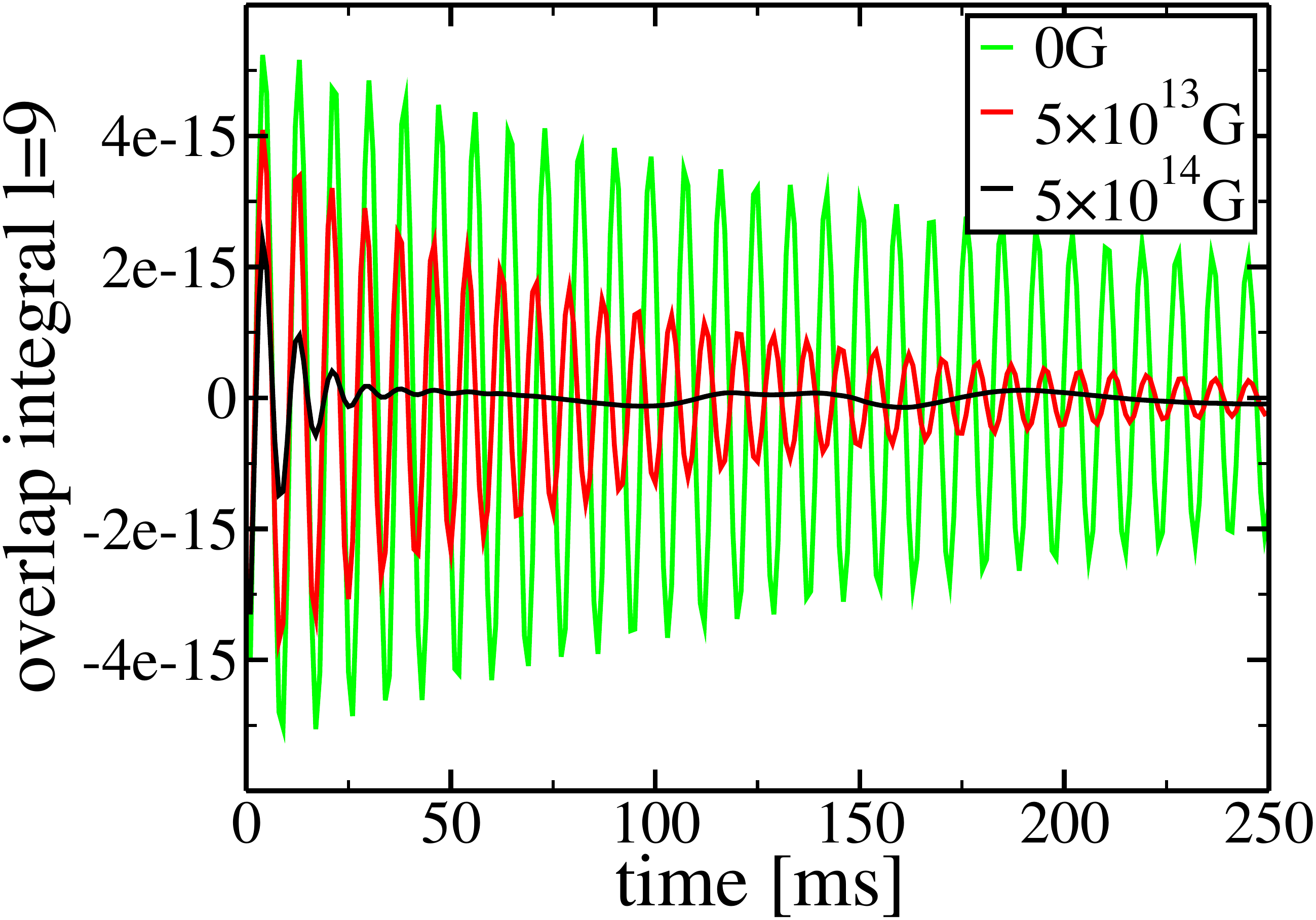}
\caption{Time evolution of overlap integrals with the eigenmodes of
  the crust. \emph{Left panel}: Damping of $l=2,3$, and $9$ initial
  perturbations due to resonant absorption of the fundamental ($n=0$)
  crustal shear mode for a magnetised model with $5\times 10^{13}\,$G
  (dots). In the corresponding unmagnetised models (solid lines) only
  numerical damping occurs which increases with the angular order $l$
  of the mode.  \emph{Middle panel}: Overlap integrals for the $l=2$
  mode, where resonant absorption of the crustal modes becomes
  stronger with increasing magnetic field strength. \emph{Right
    panel}: same as middle panel, but for $l=9$.}
\label{fig_damping}
\end{figure*}

We investigated perturbations of different radial extent encompassing
only the crust, only the core, or the whole star. Since all three
types of perturbations give qualitatively similar results, we focus on
whole star perturbations in the following, as this is the most generic
case. We first consider initial perturbations  consisting of a single 
torsional, spherical vector harmonics $l$-mode.
To investigate the damping of a single crustal mode we compute {\it
  overlap integrals} of the evolved variables with mode eigenfunctions
(the latter are found by solving the linear eigenvalue problem for the
crustal modes, see \cite{Schumaker1983,Messios2001,Sotani2007}).
Because the eigenmodes of the crust form a complete orthonormal set we
can expand any perturbation in terms of the corresponding
eigenfunctions.  The expansion factors, which provide a measure of how
strong each crustal mode contributes to the perturbation, are obtained
via the overlap integrals with the eigenfunctions. For more details on
this method see \cite{Gabler2009}. In the left panel of
Fig.\,\ref{fig_damping} we show the maximum (absolute) amplitudes of
the overlap integrals for different initial perturbations and for
simulations both without magnetic field (solid lines) and with a polar
magnetic field of $5\times10^{13}$G (dots) . In the field-free case
the lines represent the {\it numerical damping} of crustal modes due
to finite-differencing.  When a magnetic field is present, the damping
(now due to resonant absorption) increases with the magnetic field
strength.
\begin{table}
\begin{center}
\begin{tabular}{c | c c  c c}
  $l$                                  & 2      & 3    & 9     & 10   \\ \hline
  $\tau\,$[s] for $B=0$                & 5.500 & 4.620 & 0.260 & 0.170\\
  $\tau\,$[s] for $B=5\times10^{13}\,$G & 0.072 & 0.080 & 0.087 & 0.081\\
  $\tau\,$[s] for $B=10^{14}\,$G & 0.040 & 0.040 & 0.022 & 0.021\\ \hline
\end{tabular}
\end{center}
\caption{Damping timescales $\tau$ due to resonant absorption of
  crustal shear modes by the Alfv\'en continuum for various
  initial perturbation modes $l$.}
\label{tab_damping}
\end{table}
\begin{table}
\begin{center}
\begin{tabular}{c | c c c}
  EOS    &\multicolumn{3}{c}{$\tau\,$[ms] at $B=10^{14}\,$G}\\
 &                 $n=0$, $l=2$ & $n=0$, $l=3 $ &$n=0$, $l=9$\\
\hline
A+DH 1.6 	&23&23&15\\
A+NV 1.6 	&39&43&75\\
APR+DH 2.0	&32&31&17\\
APR+NV 2.0	&50&54&80\\
L+DH 1.6	&50&52&24\\
L+NV 1.6	&79&86&53\\
L+DH 2.0	&44&46&21\\
L+NV 2.0	&73&80&87\\
W+DH 1.6	&29&30&14\\
W+NV 1.6	&53&58&69\\
\hline 
\end{tabular}
\end{center}
\caption{Damping timescales $\tau$ due to resonant absorption of
  crustal shear modes by the Alfv\'en continuum for
  initial perturbation modes $l=2$, $l=3$ and $l=9$ and for different
  combinations of equations of state at $B = 10^{14}\,$G. The number 
  in the labelling of the EOS represents the mass of the neutron star 
  model in M$_\odot$ }
\label{tab_damping_eos}
\end{table}

For all modes, the timescale of resonant absorption is much shorter
than that of numerical damping (see Table \ref{tab_damping}).  After about
500\,ms, the overlap integrals no longer sample the crust
oscillations, but instead the magneto-elastic oscillations which then
dominate the evolution (see below).

The middle and right panels of Fig.\,\ref{fig_damping} show the
overlap integrals for $l=2$ and $l=9$ modes of the crust as a function
of time for different magnetic field strengths, respectively. For
$l=2$ and $B = 5\times10^{13}\,$G we find almost complete damping of
the crustal mode after $\sim 0.5\,$s. For a stronger magnetic field
($B = 5\times10^{14}\,$G) the crustal mode becomes already damped
after less than one oscillation, and only the dominant magneto-elastic
oscillations remain.  Similar statements hold for the evolution of the
$l=9$ perturbation. However, in that case it takes a few oscillations
before the crustal mode is damped, even for $B = 5\times10^{14}\,$G.

\begin{figure}
\centering
\includegraphics[width=.38\textwidth]{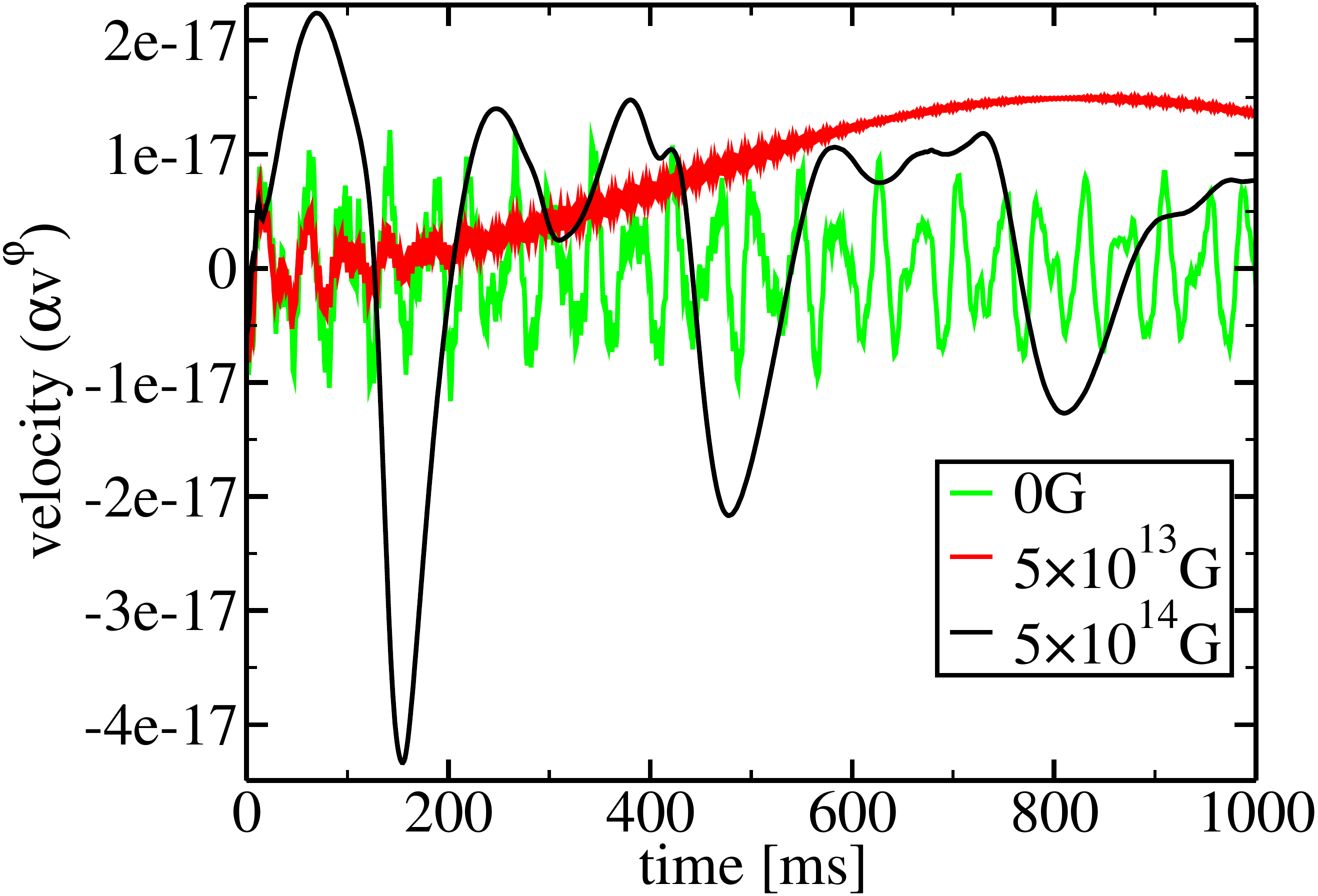}
\caption{Evolution of $\alpha v^\varphi$ in the crust at
  $\theta=\pi/4$ for initial data containing a large number of
  different perturbation modes.  For vanishing magnetic field, only
  crustal modes are excited. For $B = 5\times 10^{13}\,$G the crustal
  oscillations are strongly damped, and for a ten times stronger
  magnetic field ($B = 5\times 10^{14}\,$G) magneto-elastic
  oscillations dominate the evolution from the start.}
\label{time_evo}
\end{figure}

\begin{figure*}
\includegraphics[width=.7\textwidth]{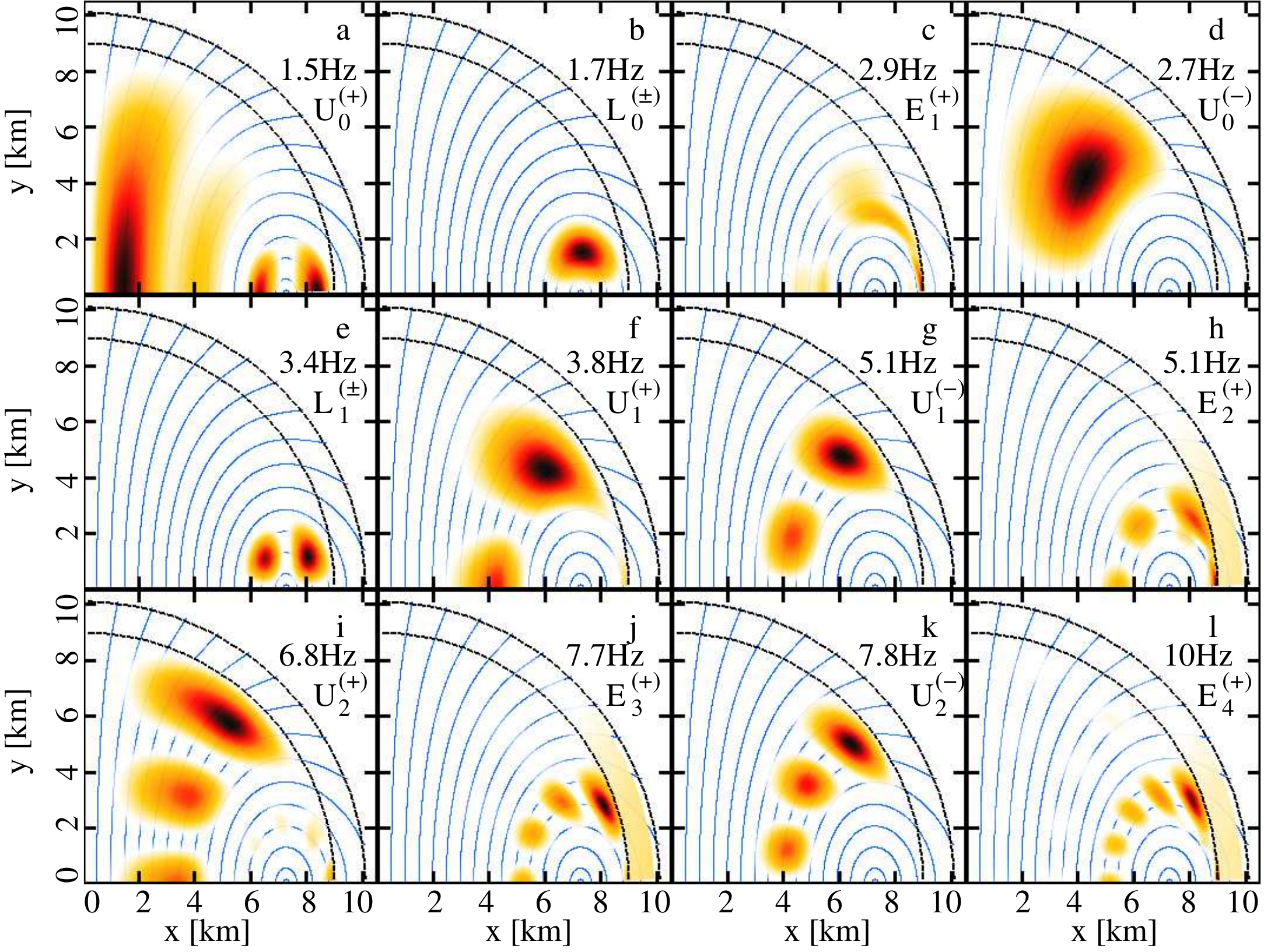}
\caption{Spatial distribution of Fourier amplitudes at several QPO
  frequencies. Three different types of QPOs are present: (i) upper
  QPOs in panels \emph{a, d, f, g, i}, and \emph{k}; (ii) lower QPOs
  in panels \emph{b} and \emph{e}; (iii) edge QPOs in panels \emph{c,
    h, j}, and \emph{l}. The two black dashed lines mark the location
  of the crust, while the blue lines represent magnetic field
  lines. The colour scale ranges from zero amplitude (white) to
  maximum amplitude (black).}
 \label{FFT_core}
 \end{figure*}

We also analysed a more general initial perturbation consisting of a
mixture of $l=2$ up to $l=10$ modes, which excites a large number of
crustal modes of different angular order $l$ and radial order $n$.
Fig.\,\ref{time_evo} shows the resulting evolution of the velocity in
the crust at $\theta=\pi/4$. For the unmagnetised model, low
frequency, fundamental ($n=0$) oscillations can easily be
distinguished from higher frequency overtones with $n \ge 1$. When
increasing the magnetic field strength to $5\times10^{13}\,$G
the lower frequency modes are completely damped after $\sim 250\,$ms,
whereas the higher frequency overtones survive for a longer time. In
the long run, a low frequency ($\sim 0.3\,$Hz) magneto-elastic
oscillation dominates. At the largest magnetic field strength shown
here, $5\times10^{14}\,$G, there is no sign of either low or high
frequency crustal modes, the evolution being completely dominated by
magneto-elastic oscillations.

Damping timescales of $n=0$, $l=\{2,3,9\}$ modes for different EOSs
are listed in Table \ref{tab_damping_eos}. For all EOSs and masses
studied here, we find significant damping of the crustal shear modes
at $B=10^{14}\,$G.  Up to the highest-order $l=9$ mode, the damping
timescales are shorter than 0.1s. From the results in Tables 1
  and 2, we deduce that even at $B=5\times10^{13}\,$G, which is up to
  two orders of magnitude weaker than assumed magnetar field
  strengths, the damping timescale is {\it shorter than 0.2s}, which is more
  than two orders of magnitude shorter than the duration of observed
  QPOs.

\vspace{-0.5cm}
%-------------------------------------------------------------------------------
\section{Long-term QPOs}

Besides the damping of crustal modes, we observe long-lasting
oscillations in the fluid core of the magnetar. These \emph{long-term
  QPOs} are identified by local maxima in Fourier space.  Let us
consider an intermediate magnetic field strength of
$4\times10^{14}\,$G, where both the magnetic field and the crust
influence the dynamics.  As in the case without crust
\citep{Sotani2008, Cerda2009} we find two different families of
long-term QPOs, as demonstrated in Fig.\,\ref{FFT_core}, which shows
the spatial distribution of Fourier amplitudes at several QPO
frequencies. The \emph{lower} QPOs ($L_n^{(\pm)}$) are located inside
the region of closed field lines, while the \emph{upper} QPOs
($U_n^{(\pm)}$) concentrate along open magnetic field lines closer to
the poles. We computed QPOs of either odd $(-)$ or even $(+)$ parity
w.r.t.\ the equatorial plane, which allows for a better identification
of QPOs of similar frequency but opposite parity.

The lower QPOs (Fig.\,\ref{FFT_core}, panels \emph{b} and \emph{e})
appear to be similar to those found for models without crust, except
that they are limited to the region of closed magnetic field lines
inside the core. The upper QPOs are influenced by the presence of the
crust in several ways. First, they are limited to the fluid region,
and become vanishingly small at the base of the crust
(Fig.\,\ref{FFT_core}, panels \emph{a, d, f, g, i}, and \emph{k}),
where they are reflected.  This behaviour is similar to that caused by
the boundary conditions in \cite{Sotani2008}, and differs from that of
the pure fluid case considered in \cite{Cerda2009}. In the latter work
the continuous traction boundary condition imposed at the surface of
the star resulted in a strong displacement there. Their different
behaviour at the base of the crust (w.r.t. the pure fluid case in
\cite{Cerda2009}) causes a rearrangement of the QPOs. Here, the lowest
frequency QPO (panel \emph{a}) is symmetric w.r.t. the equatorial
plane (even parity) - while it did not exist at all in \cite{Cerda2009} -
and the lowest-frequency QPO has odd parity.

While QPOs are located close to the symmetry axis of the field (polar
axis) in models without crust \citep{Sotani2008, Cerda2009}, they are
attached to field lines crossing the equator at around $4\,$km in our
models including the crust.  Apparently the strong coupling introduced
by the crust complicates the oscillatory behaviour of the field lines,
such that the interaction between neighbouring polar field lines
prevents the QPOs from being established.

Furthermore, we find a new family of QPOs (Fig.\,\ref{FFT_core},
panels \emph{c, h, j}, and \emph{l}) connected to the last open field
line of the fluid core, each member representing the lower-frequency
edge of an Alfv\'en continuum along the open field lines. 
QPOs at similar locations were also identified in simulations by
\cite{Colaiuda2009} without a crust.

\section{Discussion}
In this Letter we have presented the first numerical simulations of
axisymmetric, torsional, magneto-elastic oscillations in a realistic magnetar 
model, including an extended crust.

We focus on the timescales for resonant absorption of low frequency, $n=0$ 
crustal shear modes by the Alfv\'en continuum of the core and find that 
even at $B=5\times10^{13}\,$G, which is up to two orders of magnitude 
weaker than assumed magnetar field strengths, the damping timescale is
shorter than 0.2s, which is more than two orders of magnitude shorter
than the duration of observed QPOs. Furthermore, the crust EOSs NV and
DH which we have used have a very high shear modulus,
compared to a range of other proposed EOSs
\citep{Steiner2009}. Comparing the damping timescales obtained here
for the NV and DH EOSs, we find that, at a given magnetic field
strength, a lower shear modulus results in a shorter damping timescale
due to resonant absorption. Thus, for lower values of the shear
  modulus than considered here, the deduced damping timescales would
  be even shorter than our current findings.  In addition, no
significant excitation of crustal modes by the Alfv\'en continuum was
observed at the end of our simulations, which reached up to several
seconds. The above results do not leave much room for shear modes of
the crust to be able to sustain oscillations that lead to significant
modulations in the X-ray tail of giant SGR bursts lasting for several
tens of seconds, when the magnetic field is global. The case of 
magnetic field configurations confined to the crust, which may be realised 
if the core is a type I superconductor, requires further investigation.

In contrast to the shear modes, the Alfv\'en QPO model (see
\cite{Levin2007, Sotani2008, Cerda2009, Colaiuda2009}) has several
attractive features, matching to observed frequencies for a dipole
field strength up to several times $10^{15}\,$G and explaining the
observed integer ratios for the frequencies of QPOs.  Here we find
that for magnetic field strengths less than about $10^{15}\,$G,
 the Alfv\'en QPOs are mostly confined to the fluid core. For EOSs
with a smaller shear modulus, this could reduce to several times
$10^{14}\,$G. Above this {\it minimum field strength}, Alfv\'en
oscillations completely dominate in the crust and the resulting QPOs
have a maximum amplitude at the surface of the star. It is thus
likely, that the Alfv\'en QPO model can operate efficiently only above
such a minimum field strength. In that case, the possible
  strength of magnetar magnetic fields is limited to a narrow range,
  between the minimum strength discussed here and the upper bound
  determined in \cite{Cerda2009}. Furthermore, the empirical formulas
for the QPO frequencies presented in \cite{Cerda2009} could be used to
directly constrain the strength of the magnetic field and the
compactness of the star.

A more extended investigation of the minimum magnetic field strength
for the Alfv\'en QPO model will appear in \cite{Gabler2010}.  We are
also planning to extend our model by taking into account additional
effects, such as different magnetic field topologies (see also
\cite{Sotani2008b}), the coupling of the interior dynamics to an
external magnetosphere, and the effect of field-line entanglement in
the core.

%===============================================================================
\section*{Acknowledgements}
We are grateful to the anonymous referee for useful comments which helped to improve
the final version of this letter. This work was supported by the 
Collaborative Research Center on Gravitational
Wave Astronomy of the Deutsche Forschungsgemeinschaft (DFG
SFB/Transregio 7), the Spanish {\it Ministerio de Educaci\'on y
  Ciencia} (AYA 2007-67626-C03-01), the ESF grant COMPSTAR  
and a DAAD exchange grant. Computing time was provided by the {\it Servicio de
 Inform\'atica de la Universidad de Valencia}.

% Create the reference section using BibTeX:
\bibliographystyle{mn2e}
\bibliography{magnetar}

\end{document}